\documentclass[10pt]{iopart}
\usepackage{iopams}
\usepackage{amsmath_iop}
\usepackage{graphicx}
\usepackage{dcolumn}
\usepackage{bm}
\usepackage{color}
\usepackage{CJK}
\usepackage{url}

\def\ve#1{{\bm{#1}}}

\def\urm#1{\scriptstyle{\text{\textrm{\textmd{\textup{#1}}}}}}

\let\temp\epsilon
\let\epsilon\varepsilon
\let\varepsilon\temp
\let\temp\relax
\def\defeq{\mathrel{:=}}

\begin{document}
%
\begin{CJK*}{UTF8}{}
  \title[Improvement of Functionals in DFT by IKS \& DFPT]
  {Improvement of functionals in density functional theory 
    by the inverse Kohn--Sham method and density functional perturbation theory
  }
  \author{
    Tomoya Naito (\CJKfamily{min}{内藤智也})$ {}^{1, \, 2} $,
    Daisuke Ohashi (\CJKfamily{min}{大橋大介})$ {}^{1, \, 2} $,
    and
    Haozhao Liang (\CJKfamily{gbsn}{梁豪兆})$ {}^{2, \, 1} $}
  \address{$ {}^1 $ Department of Physics, Graduate School of Science, The University of Tokyo,
    Tokyo 113-0033, Japan}
  \address{$ {}^2 $ RIKEN Nishina Center, Wako 351-0198, Japan}
  \ead{haozhao.liang@riken.jp}
  \date{\today}
  \begin{abstract}
    We propose a way to improve energy density functionals (EDFs) in the density functional theory based on the combination of the inverse Kohn--Sham method and the density functional perturbation theory.
    Difference between the known EDF and the exact one is treated as the first-order perturbation.
    As benchmark calculations, we reproduce the theoretical exchange and correlation functionals in the local density approximation.
    Systems of noble-gas atoms are used for benchmark calculations,
    and the ground-state energies and densities, as well as the functionals, are reproduced with good accuracies.
  \end{abstract}
  \submitto{\jpb}
  \maketitle
\end{CJK*}
\section{Introduction}
\label{sec:intr}
\par
Density functional theory (DFT) is one of the most successful approaches to
the calculation of the ground-state properties of the quantum many-body problems, 
such as atoms, molecules, solids
\cite{PhysRev.136.B864,PhysRev.140.A1133,RevModPhys.71.1253,RevModPhys.87.897},
and nuclear systems \cite{Bender2003Rev.Mod.Phys.75_121,RevModPhys.88.045004}.
Since for the same level of accuracy the numerical cost of DFT is much less than those of other quantum many-body methods \cite{Engel2011_Springer-Verlag},
such as 
the quantum Monte Carlo method
\cite{
  McMillan1965Phys.Rev.138_A442,
  Ceperley1977Phys.Rev.B16_3081,
  Ceperley1980Phys.Rev.Lett.45_566,
  Needs2010J.Phys.Condens.Matter22_023201,
  Booth2013Nature493_365},
many-body perturbation theory
\cite{
  Moller1934Phys.Rev.46_618,
  Hybertsen1986Phys.Rev.B34_5390,
  Kuwahara2016Phys.Rev.B94_121116},
configuration interaction method
\cite{
  Pople1976Int.J.QuantumChem.10_1,
  Pople1977Int.J.QuantumChem.12_149,
  Pople1999Rev.Mod.Phys.71_1267},
coupled-cluster method
\cite{
  Coester1958Nucl.Phys.7_421,
  Coester1960Nucl.Phys.17_477,
  Cizek1966J.Chem.Phys.45_4256,
  Cizek1971Int.J.QuantumChem.5_359,
  Kuemmel1978Phys.Rep.36_1},
and transcorrelated method
\cite{
  Boys1969Proc.R.Soc.A309_209,
  Umezawa2003J.Chem.Phys.119_10015,
  Tsuneyuki2008Prog.Theor.Phys.Suppl.176_134}.
Thus, DFT is applicable to larger-scale calculations
\cite{Iwata2010J.Comput.Phys.229_2339,Soler2002J.Phys.Condens.Matter14_2745,Ozaki2006Phys.Rev.B74_245101}.
Furthermore, in principle, the DFT gives the exact ground-state density $ \rho_{\urm{gs}} $ and energy $ E_{\urm{gs}} $:
\begin{equation}
  \label{eq:GSenergy_def}
  E_{\urm{gs}}
  =
  T_0 \left[ \rho_{\urm{gs}} \right]
  +
  \int
  V_{\urm{ext}}
  \left( \ve{r} \right) \,
  \rho_{\urm{gs}} \left( \ve{r} \right) \,
  d \ve{r}
  +
  E_{\urm{H}} \left[ \rho_{\urm{gs}} \right]
  +
  E_{\urm{xc}} \left[ \rho_{\urm{gs}} \right],
\end{equation}
where $ T_0 $ is the Kohn--Sham (KS) kinetic energy,
$ V_{\urm{ext}} $ is the external field,
and $ E_{\urm{H}} \left[ \rho \right] $ and
$ E_{\urm{xc}} \left[ \rho \right] $ are the Hartree and exchange-correlation energy density functionals (EDFs), respectively \cite{PhysRev.136.B864,PhysRev.140.A1133}.
However, in practice, the accuracy of the DFT calculation depends on the accuracy of the approximations for $ E_{\urm{xc}} \left[ \rho \right] $,
as it is unknown.
\par
In electron systems, many approximations for $ E_{\urm{xc}} \left[ \rho \right] $ have been proposed from first principles, i.e.,~non-empirically.
The widely used ones are the local density approximation (LDA) \cite{Proc.Camb.Phil.Soc.26.376,Can.J.Phys.58_1200--1211,PhysRevB.23.5048,PhysRevB.45.13244}
and generalized gradient approximation (GGA) \cite{PhysRevA.38.3098,PhysRevB.46.6671,PhysRevLett.77.3865,PhysRevLett.100.136406}.
In the LDA,
$ E_{\urm{xc}} \left[ \rho \right] $ is approximated as a functional of local density $ \rho \left( \ve{r} \right) $,
whereas in the GGA, $ E_{\urm{xc}} \left[ \rho \right] $ is approximated as a functional of $ \rho \left( \ve{r} \right) $ and its gradient $ \left| \nabla \rho \left( \ve{r} \right) \right| $.
Approximations beyond the GGA have also been developed,
such as hybrid functionals,
and they are being applied extensively \cite{doi:10.1063/1.1390175}, in particular, in quantum chemistry.
Note that even in the GGA, $ E_{\urm{xc}} \left[ \rho \right]$ is constructed based on the LDA one.
Thus, the exchange interaction is not fully included, and hence
the physics of localized electrons
(including materials containing $ d $- and $ f $-electrons)
is not captured accurately with semi-local functionals.
In order to avoid these problems, phenomenological methods,
such as the $ \mbox{LDA} + U $ \cite{Anisimov1991Phys.Rev.B44_943,Liechtenstein1995Phys.Rev.B52_R5467,Petukhov2003Phys.Rev.B67_153106},
$ \mbox{LDA} + \mbox{DMFT} $ \cite{Anisimov1997J.Phys.Condens.Matter9_7359},
exact-exchange \cite{Goerling1996Phys.Rev.B53_7024,Staedele1997Phys.Rev.Lett.79_2089} methods,
and hybrid functionals \cite{Becke1993J.Chem.Phys.98_5648,Perdew1996J.Chem.Phys.105_9982,Adamo1998J.Chem.Phys.108_664,Heyd2003J.Chem.Phys.118_8207}
have been proposed.
Functionals with the long-range correction due to the van der Waals interaction \cite{Iikura2001J.Chem.Phys.115_3540}
and semi-empirical functionals \cite{Chai2008TheJournalofChemicalPhysics128_084106,Peverati2012J.Phys.Chem.Lett.3_117}
have also been discussed for a long time.
\par
In contrast, in nuclear systems, the exact form of the interaction in the vacuum between nucleons is still under discussion
\cite{PhysRevC.51.38,PhysRevLett.99.022001,Aoki2012Prog.Theor.Exp.Phys.2012_01A105,Hatsuda1994Phys.Rep.247_221,Holt2016Phys.Rep.621_2,doi:10.1143/PTP.17.360,PhysRevC.64.014001}. 
Even if the exact form of the interaction in the vacuum were known,
the nuclear interaction in the medium is different due to its highly non-perturbative property \cite{PhysRev.95.217,Goldstone1957Proc.R.Soc.A239_267}.
Therefore, it is still difficult to derive the Hartree-exchange-correlation EDF
$ E_{\urm{Hxc}} \left[ \rho \right] = E_{\urm{H}} \left[ \rho \right] + E_{\urm{xc}} \left[ \rho \right] $
from first principles,
although the Hartree--Fock calculation from the interaction in the vacuum has been discussed for a long time
\cite{Coester1970Phys.Rev.C1_769,RevModPhys.39.719,Shen2016Chin.Phys.Lett.33_102103,Shen2017Phys.Rev.C96_014316,Shen2018Phys.Rev.C97_054312}.
Thus, $ E_{\urm{Hxc}} \left[ \rho \right] $ for the nuclear interaction is treated phenomenologically
\cite{PhysRevC.5.626,PhysRevC.21.1568,Long2006Phys.Lett.B640_150,Meng2006Prog.Part.Nucl.Phys.57_470},
with fitting parameters to experimental data \cite{Bender2003Rev.Mod.Phys.75_121}.
Since the fitting parameters are usually determined from the experimental data of several stable nuclei,
different parameter sets can give totally different results for exotic nuclei \cite{PhysRevC.61.034313}.
Comparisons between parameter sets are still being discussed \cite{PhysRevC.58.2796,Roca-Maza2018Prog.Part.Nucl.Phys.101_96}.
\par
Hence,
the derivation or construction of accurate EDFs is one of the primary goals in DFT for both electron and nuclear systems,
whose strategies are, however, under debate \cite{Medvedev2017Science355_49,Kepp2017Science356_496}.
Recently, a new microscopic way to derive EDFs based on the functional renormalization group was suggested \cite{Liang2018Phys.Lett.B779_436,Yokota2019Phys.Rev.C99_024302,Yokota2019Prog.Theor.Exp.Phys.2019_011D01,Yokota2019Phys.Rev.B99_115106},
while it is not ready for realistic systems yet.
\par
As an alternative way to improve EDFs, the inverse approach of DFT, the so-called inverse Kohn--Sham (IKS) method, was proposed in Refs.~\cite{PhysRevA.47.R1591,doi:10.1063/1.465093}.
As mentioned in Ref.~\cite{Kohn1999Rev.Mod.Phys.71_1253},
the KS potential $ V_{\urm{KS}} \left( \ve{r} \right) $ is unique concerning the system,
and
$ V_{\urm{KS}} \left( \ve{r} \right) $ is calculated from the given ground-state density $ \rho_{\urm{gs}} \left( \ve{r} \right) $ in the IKS.
The information provided by the IKS,
such as the single-particle energies $ \epsilon_i $,
is expected to be valuable for improving the accuracy of EDFs \cite{PhysRevA.29.2322,PhysRevA.33.804,PhysRevA.50.3827}.
Although with such an expectation,
the actual way for using IKS has not been pointed out explicitly,
only the improvement of the numerical methods of IKS has been discussed \cite{Jensen2018Int.J.QuantumChem.118_e25425}.
Since the density $ \rho \left( \ve{r} \right) $ is usually the quantity which one interests in,
once $ \rho \left( \ve{r} \right) $ of the system is known, 
tasks of the calculation for the system are almost achieved.
\par
Nevertheless, improvement of the EDFs by using the IKS is promising
since the EDF is, in principle, unique for all the electron systems,
while the density of some systems, for instance, atoms and light molecules, can be determined 
from experiments or high-accuracy calculations,
such as the coupled-cluster and the configuration interaction methods.
\par
Moreover, in nuclear systems,
it is known that the effective nuclear force, i.e.,~the nuclear force in medium, is substantially different from the bare nuclear force \cite{PhysRev.95.217}
and its details are still unknown.
Therefore, the nuclear EDFs are derived from the fitting to the experimental data
\cite{Bender2003Rev.Mod.Phys.75_121,RevModPhys.88.045004}.
In contrast,
for the light or medium-heavy nuclei,
the many-body methods beginning from the bare nuclear force,
so-called \textit{ab initio} methods,
such as
the no-core shell model \cite{Navratil2007Phys.Rev.Lett.99_042501,Barrett2013Prog.Part.Nucl.Phys.69_131},
the self-consistent Green's function method \cite{Dickhoff2004Prog.Part.Nucl.Phys.52_377},
the lattice chiral effective field theory \cite{Borasoy2008Eur.Phys.J.A35_343,Epelbaum2009Eur.Phys.J.A41_125,Epelbaum2010Phys.Rev.Lett.104_142501},
the in-medium similarity renormalization group \cite{Tsukiyama2011Phys.Rev.Lett.106_222502,Tsukiyama2012Phys.Rev.C85_061304,Hergert2016Phys.Scr.92_023002,Hergert2016Phys.Rep.621_165},
and the Brueckner Hartree--Fock method 
\cite{Coester1970Phys.Rev.C1_769,RevModPhys.39.719,Shen2016Chin.Phys.Lett.33_102103,Shen2017Phys.Rev.C96_014316,Shen2018Phys.Rev.C97_054312,Tong2018Phys.Rev.C98_054302,Zhang2018Phys.Rev.C98_064306,Shen2019_arXiv_1904.04977},
are available.
These \textit{ab initio} methods provide accurate density distributions for light nuclei,
but meanwhile, their calculations for heavy nuclei are impossible even in the visible future.
Therefore, how to make the best use of these \textit{ab initio} methods for developing nuclear DFT is a worldwide hot topic in the nuclear community
\cite{Shen2019_arXiv_1904.04977,Drut2010Prog.Part.Nucl.Phys.64_120,Shen2019Phys.Rev.C99_034322}.
\par
The density functional perturbation theory (DFPT) has been developed for decades.
The DFPT was originally developed for the calculations of phonon and response properties from the first principle \cite{PhysRevLett.58.1861,PhysRevA.52.1096,Gonze1989Phys.Rev.B39_13120,RevModPhys.73.515}.
Properties of solids are determined from phonons as well as electrons \cite{Ashcroft2016_Cengage},
but the original DFT is applicable only for the electronic structure.
In the DFPT, the displacement of the external field $ V_{\urm{ext}} $
caused by the displacement of the nuclei or external electric field
is treated by a combination of the perturbation and linear response theories \cite{RevModPhys.73.515}.
In principle, each order of the DFPT
can be derived systematically
\cite{Gonze1995Phys.Rev.A52_1096},
and so far, the DFPT up to the third order has been discussed \cite{Gonze1989Phys.Rev.B39_13120}.
The DFPT has also been applied to derive the EDF \cite{Goerling1993Phys.Rev.B47_13105,Goerling1994Phys.Rev.A50_196}.
\par
In order to attack the open question about the practical use of the IKS, 
for the first time, a new strategy based on the combination of the DFPT and the IKS, 
the so-called IKS-DFPT, is proposed in this paper.
The first-order DFPT, which is also called the Hellmann--Feynman theorem \cite{Feynman1939Phys.Rev.56_340}, is used in this paper.
The known functional is improved by using the IKS-DFPT.
The ground-state energy derived by two methods:
One is a combination of the original and inverse Kohn--Sham scheme,
and the other is the DFPT.
In the latter way, the difference between the ``exact'' functional and the known functional is assumed to be small enough,
and the difference is treated as the perturbation.
Note that the DFPT is used as the perturbation for the functional itself instead of the external potential.
\par
As benchmark calculations, we verify this method by reproducing both the LDA exchange functional \cite{Proc.Camb.Phil.Soc.26.376} and the LDA correlation functional \cite{PhysRevB.23.5048}.
The iteration of IKS-DFPT is also discussed.
\par
This paper is organized as follows:
First, the theoretical framework of the IKS-DFPT is given in Sec.~\ref{sec:theo}.
Then, the benchmark calculations and discussion of their results are shown in Sec.~\ref{sec:calc}.
Finally, Sec.~\ref{sec:conc} is devoted to the conclusion and perspectives.
%
\section{Theoretical Framework}
\label{sec:theo}
\subsection{Formalism of IKS-DFPT}
\par
In the DFT, the ground-state density $ \rho_{\urm{gs}} \left( \ve{r} \right) $
and energy $ E_{\urm{gs}} $ of an $ N $-particle system are obtained by solving the KS equations self-consistently
\begin{equation}
  \label{eq:KSeq}
  \left[
    -\frac{\hbar^{2}}{2m}
    \nabla^{2}
    +
    V_{\urm{KS}} \left( \ve{r} \right)
  \right]
  \psi_i \left( \ve{r} \right)
  =
  \epsilon_i
  \psi_i \left( \ve{r} \right),
\end{equation}
where $ m $ is the mass of particles,
$ \psi_i \left( \ve{r} \right) $ and $ \epsilon_i $ are the single-particle orbitals and energies,
respectively,
and $ \rho_{\urm{gs}} \left( \ve{r}\right) = 
\sum_{i=1}^N
\left|
  \psi_i \left( \ve{r} \right)
\right|^{2} $.
Here, $ V_{\urm{KS}} \left( \ve{r} \right) $ is the KS effective potential defined as
\begin{equation}
  V_{\urm{KS}} \left( \ve{r} \right)
  =
  V_{\urm{ext}} \left( \ve{r} \right)
  +
  \frac{\delta E_{\urm{Hxc}} \left[ \rho_{\urm{gs}} \right]}{\delta \rho \left( \ve{r} \right)}.
\end{equation}
\par
The IKS provides this KS effective potential $ V_{\urm{KS}} $ for each system from the ground-state density $ \rho_{\urm{gs}} $.
In the novel method IKS-DFPT,
the conventional Hartree-exchange-correlation functional $ \tilde{E}_{\urm{Hxc}} $,
such as the PZ81 \cite{PhysRevB.23.5048} and PBE \cite{Perdew1996Phys.Rev.Lett.77_3865,Perdew1996Phys.Rev.B54_16533} functionals,
will be improved by using the IKS.
\par
Here,
$ \tilde{E}_{\urm{Hxc}} \left[ \rho \right] $ is assumed to be close enough to the exact one
$ E^{\urm{exact}}_{\urm{Hxc}} \left[ \rho \right]$,
since $ \tilde{E}_{\urm{Hxc}} \left[ \rho \right] $ is known to work well.
Hence, the difference between
$ E_{\urm{Hxc}}^{\urm{exact}} $ and $ \tilde{E}_{\urm{Hxc}} $ 
is treated as a perturbation.
If the difference is not small enough to be treated as the perturbation,
the final results would be unreasonable.
\par
In this paper, the first-order perturbation theory is used for the treatment of the difference between $ E_{\urm{Hxc}}^{\urm{exact}} $ and $ \tilde{E}_{\urm{Hxc}} $ as
\begin{equation}
  \label{eq:idea_Hxc}
  E^{\urm{exact}}_{\urm{Hxc}} \left[ \rho \right]
  =
  \tilde{E}_{\urm{Hxc}} \left[ \rho \right]
  +
  \lambda E^{(1)}_{\urm{Hxc}} \left[ \rho \right]
  +
  O \left( \lambda^{2} \right),
\end{equation}
with a small parameter $ \lambda $.
Then, the exact single-particle orbitals $ \psi^{\urm{exact}}_i \left( \ve{r} \right) $, 
ground-state density $ \rho^{\urm{exact}}_{\urm{gs}} \left( \ve{r} \right) $,
and energy $ E^{\urm{exact}}_{\urm{gs}} $ are also expanded perturbatively:
\numparts
\begin{align}
  \psi^{\urm{exact}}_i \left( \ve{r} \right)
  & =
    \tilde{\psi}_i \left( \ve{r} \right)
    +
    \lambda \psi^{(1)}_i \left( \ve{r} \right)
    +
    O \left( \lambda^{2} \right),
    \label{eq:idea_psi} \\
  \rho^{\urm{exact}}_{\urm{gs}} \left( \ve{r} \right)
  & =
    \tilde{\rho}_{\urm{gs}} \left( \ve{r} \right)
    +
    \lambda \rho_{\urm{gs}}^{(1)} \left( \ve{r} \right)
    +
    O \left( \lambda^{2} \right),
    \label{eq:idea_rho} \\
  E^{\urm{exact}}_{\urm{gs}}
  & =
    \tilde{E}_{\urm{gs}} +\lambda E_{\urm{gs}}^{(1)}
    +
    O \left( \lambda^{2} \right),
    \label{eq:idea_GSenergy}
\end{align}
\endnumparts
where quantities shown with the tilde are given by $ \tilde{E}_{\urm{Hxc}} $.
According to these definitions,
\begin{equation}
  \label{eq:dens_1}
  \rho_{\urm{gs}}^{(1)} \left( \ve{r} \right)
  =
  \sum_{i = 1}^N
  \left[
    \psi_i^{(1)*} \left( \ve{r} \right) \,
    \psi_i \left( \ve{r} \right)
    + 
    \psi_i^* \left( \ve{r} \right) \,
    \psi_i^{(1)} \left( \ve{r} \right)
  \right]
\end{equation}
is hold.
Also, the first-order perturbation term $ \psi_i^{(1)} $ is assumed to be orthogonal to $ \tilde{\psi}_i $;
\begin{equation}
  \int
  \tilde{\psi}_i^* \left( \ve{r} \right) \,
  \psi_i^{(1)} \left( \ve{r} \right) \,
  d \ve{r}
  = 0 .
\end{equation}
\par
The perturbation is assumed not to affect the external field,
i.e.,~$ V^{\urm{exact}}_{\urm{ext}} \left( \ve{r} \right) = \tilde{V}_{\urm{ext}} \left( \ve{r} \right) $.
Moreover, $ \rho^{\urm{exact}}_{\urm{gs}} \left( \ve{r} \right) $ is assumed to be given,
and thus $ \psi^{\urm{exact}}_i \left( \ve{r} \right) $ are calculated from the IKS.
\par
Under these assumptions, we calculate $ E^{\urm{exact}}_{\urm{gs}} $ in two different ways.
One way is based on the first-order DFPT, and the other way is based on the IKS and KS equation.
In the former way, substitution of Eqs.~\eqref{eq:idea_Hxc}, \eqref{eq:idea_psi}, and \eqref{eq:idea_rho} into Eq.~\eqref{eq:GSenergy_def} gives
\fl
\begin{align}
  E_{\urm{gs}}^{\urm{exact}}
  = & \,
      T_0 \left[ \rho_{\urm{gs}}^{\urm{exact}} \right]
      +
      \int
      V_{\urm{ext}} \left( \ve{r} \right)
      \rho_{\urm{gs}}^{\urm{exact}} \left( \ve{r} \right) \,
      d \ve{r} 
      +
      E_{\urm{Hxc}}^{\urm{exact}} \left[ \rho_{\urm{gs}}^{\urm{exact}} \right]
      \notag \\
  = & \,
      T_0 \left[ \rho_{\urm{gs}}^{\urm{exact}} \right]
      +
      \int
      V_{\urm{ext}} \left( \ve{r} \right) \,
      \left[
      \tilde{\rho}_{\urm{gs}} \left( \ve{r} \right)
      +
      \lambda \rho_{\urm{gs}}^{(1)} \left( \ve{r} \right) 
      \right]
      d \ve{r} 
      \notag \\
    & +
      E_{\urm{Hxc}}^{\urm{exact}} \left[ \tilde{\rho}_{\urm{gs}} + \lambda \rho_{\urm{gs}}^{(1)} \right]
      +
      O \left( \lambda^2 \right)
      \notag \\
  = & \,
      T_0 \left[ \rho_{\urm{gs}}^{\urm{exact}} \right]
      +
      \int
      V_{\urm{ext}} \left( \ve{r} \right) \,
      \tilde{\rho}_{\urm{gs}} \left( \ve{r} \right) \,
      d \ve{r}
      +
      E_{\urm{Hxc}}^{\urm{exact}} \left[ \tilde{\rho}_{\urm{gs}} \right] \notag \\
    & +
      \lambda
      \int
      V_{\urm{ext}} \left( \ve{r} \right) \,
      \rho_{\urm{gs}}^{(1)} \left( \ve{r} \right) \,
      d \ve{r}
      +
      \lambda
      \int
      \frac{\delta E_{\urm{Hxc}}^{\urm{exact}} \left[ \tilde{\rho}_{\urm{gs}} \right]}{\delta \rho \left( \ve{r} \right)}
      \rho_{\urm{gs}}^{(1)} \left( \ve{r} \right) \,
      d \ve{r} 
      +
      O \left( \lambda^2 \right)
      \notag \\
  = & \,
      T_0 \left[ \rho_{\urm{gs}}^{\urm{exact}} \right]
      +
      \int
      V_{\urm{ext}} \left( \ve{r} \right) \,
      \tilde{\rho}_{\urm{gs}} \left( \ve{r} \right) \,
      d \ve{r}
      +
      \tilde{E}_{\urm{Hxc}} \left[ \tilde{\rho}_{\urm{gs}} \right]
      +
      \lambda
      E_{\urm{Hxc}}^{(1)} \left[ \tilde{\rho}_{\urm{gs}} \right]
      \notag \\
    & +
      \lambda
      \int
      V_{\urm{ext}} \left( \ve{r} \right) \,
      \rho_{\urm{gs}}^{(1)} \left( \ve{r} \right) \,
      d \ve{r}
      +
      \lambda
      \int
      \frac{\delta \tilde{E}_{\urm{Hxc}} \left[ \tilde{\rho}_{\urm{gs}} \right]}{\delta \rho \left( \ve{r} \right)}
      \rho_{\urm{gs}}^{(1)} \left( \ve{r} \right) \,
      d \ve{r} 
      +
      O \left( \lambda^2 \right),
      \label{eq:gs_1}
\end{align}
where the Taylor expansion for general functionals $ F \left[ \rho \right] $ 
\begin{align}
  & F \left[ \rho_0 + \delta \rho \right]
    \notag \\
  & =
    F \left[ \rho \right]
    +
    \int
    \frac{\delta F \left[ \rho_0 \right]}{\delta \rho \left( \ve{r} \right)}
    \delta \rho \left( \ve{r} \right) \,
    d \ve{r} 
    +
    \frac{1}{2}
    \iint
    \frac{\delta^2 F \left[ \rho_0 \right]}{\delta \rho \left( \ve{r} \right) \, \delta \rho \left( \ve{r}' \right)}
    \delta \rho \left( \ve{r} \right) \, 
    \delta \rho \left( \ve{r}' \right) \, 
    d \ve{r} \, d \ve{r}'
    +
    \cdots
\end{align}
is used.
Here, the kinetic term $ T_0 $ satisfies
\begin{align}
  T_0 \left[ \rho_{\urm{gs}}^{\urm{exact}} \right]
  = & \,
      \sum_{i = 1}^N
      \int
      \psi_i^* \left( \ve{r} \right) \,
      \left(
      - \frac{\hbar^2}{2m} \nabla^2
      \right)
      \psi_i \left( \ve{r} \right)
      \, d \ve{r}
      \notag \\
  = & \,
      - \frac{\hbar^2}{2m}
      \sum_{i = 1}^N
      \int
      \left(
      \tilde{\psi}_i^* \left( \ve{r} \right) + \lambda \psi_i^{(1)*} \left( \ve{r} \right)
      \right)
      \nabla^2
      \left(
      \tilde{\psi}_i \left( \ve{r} \right) + \lambda \psi_i^{(1)} \left( \ve{r} \right)
      \right)
      \, d \ve{r}
      + O \left( \lambda^2 \right)
      \notag \\
  = & \, 
      - \frac{\hbar^2}{2m}
      \sum_{i = 1}^N
      \left[
      \int
      \tilde{\psi}_i^* \left( \ve{r} \right) \,
      \nabla^2
      \tilde{\psi}_i \left( \ve{r} \right) 
      \, d \ve{r}
      \right.
      \notag \\
    & \left.
      +
      \lambda
      \int
      \left\{
      \tilde{\psi}_i^* \left( \ve{r} \right) \,
      \nabla^2
      \psi_i^{(1)} \left( \ve{r} \right) 
      +
      \psi_i^{(1)*} \left( \ve{r} \right) \,
      \nabla^2
      \tilde{\psi}_i \left( \ve{r} \right) 
      \right\}
      \, d \ve{r}
      \right]
      +
      O \left( \lambda^2 \right)
      \notag \\
  = & \,
      T_0 \left[ \tilde{\rho}_{\urm{gs}} \right]
      \notag \\
    & - 
      \frac{\lambda \hbar^2}{2m}
      \sum_{i = 1}^N
      \int
      \left\{
      \tilde{\psi}_i^* \left( \ve{r} \right) \,
      \nabla^2
      \psi_i^{(1)} \left( \ve{r} \right) 
      +
      \psi_i^{(1)*} \left( \ve{r} \right) \,
      \nabla^2
      \tilde{\psi}_i \left( \ve{r} \right) 
      \right\}
      \, d \ve{r}
      +
      O \left( \lambda^2 \right)
      \label{eq:T0_exp}.
\end{align}
Combining Eqs.~\eqref{eq:dens_1}, \eqref{eq:gs_1}, and \eqref{eq:T0_exp}, we get
\begin{align}
  \tilde{E}_{\urm{gs}}
  = & \,
      T_0 \left[ \tilde{\rho}_{\urm{gs}} \right]
      +
      \int	
      V_{\urm{ext}} \left( \ve{r} \right) \,
      \tilde{\rho}_{\urm{gs}} \left( \ve{r} \right) \,
      d \ve{r}
      +
      \tilde{E}_{\urm{Hxc}} \left[ \tilde{\rho}_{\urm{gs}} \right], \\
  E_{\urm{gs}}^{(1)}
  = & \,
      - 
      \frac{\hbar^2}{2m}
      \sum_{i = 1}^N
      \int
      \left\{
      \tilde{\psi}_i^* \left( \ve{r} \right) \,
      \nabla^2
      \psi_i^{(1)} \left( \ve{r} \right) 
      +
      \psi_i^{(1)*} \left( \ve{r} \right) \,
      \nabla^2
      \tilde{\psi}_i \left( \ve{r} \right) 
      \right\}
      \, d \ve{r}
      \notag \\
    & +
      \int
      V_{\urm{ext}} \left( \ve{r} \right) \,
      \rho_{\urm{gs}}^{(1)} \left( \ve{r} \right) \,
      d \ve{r}
      +
      \int
      \frac{\delta \tilde{E}_{\urm{Hxc}} \left[ \tilde{\rho}_{\urm{gs}} \right]}{\delta \rho \left( \ve{r} \right)}
      \rho_{\urm{gs}}^{(1)} \left( \ve{r} \right) \,
      d \ve{r}
      +
      E_{\urm{Hxc}}^{(1)} \left[ \tilde{\rho}_{\urm{gs}} \right]
      \notag \\
  = & \,
      - 
      \frac{\hbar^2}{2m}
      \sum_{i = 1}^N
      \int
      \left\{
      \tilde{\psi}_i^* \left( \ve{r} \right) \,
      \nabla^2
      \psi_i^{(1)} \left( \ve{r} \right) 
      +
      \psi_i^{(1)*} \left( \ve{r} \right) \,
      \nabla^2
      \tilde{\psi}_i \left( \ve{r} \right) 
      \right\}
      \, d \ve{r}
      \notag \\
    & +
      \sum_{i = 1}^N
      \int
      V_{\urm{ext}} \left( \ve{r} \right) \,
      \left\{
      \psi_i^{(1)*} \left( \ve{r} \right) \,
      \tilde{\psi}_i \left( \ve{r} \right)
      +
      \tilde{\psi}_i^* \left( \ve{r} \right) \,
      \psi_i^{(1)} \left( \ve{r} \right)
      \right\}
      \, d \ve{r}
      \notag \\
    & + \sum_{i = 1}^N
      \int
      \frac{\delta \tilde{E}_{\urm{Hxc}} \left[ \tilde{\rho}_{\urm{gs}} \right]}{\delta \rho \left( \ve{r} \right)}
      \left\{
      \psi_i^{(1)*} \left( \ve{r} \right) \,
      \tilde{\psi}_i \left( \ve{r} \right)
      +
      \tilde{\psi}_i^* \left( \ve{r} \right) \,
      \psi_i^{(1)} \left( \ve{r} \right)
      \right\}
      \, d \ve{r}
      +
      E_{\urm{Hxc}}^{(1)} \left[ \tilde{\rho}_{\urm{gs}} \right]
      \notag \\
  = & \,
      \sum_{i = 1}^N
      \int
      \tilde{\psi}_i^* \left( \ve{r} \right) \,
      \left\{
      - 
      \frac{\hbar^2}{2m}
      \nabla^2
      +
      V_{\urm{ext}} \left( \ve{r} \right)
      +
      \frac{\delta \tilde{E}_{\urm{Hxc}} \left[ \tilde{\rho}_{\urm{gs}} \right]}{\delta \rho \left( \ve{r} \right)}
      \right\}
      \psi_i^{(1)} \left( \ve{r} \right) 
      \, d \ve{r}
      \notag \\
    & +
      \sum_{i = 1}^N
      \int
      \psi_i^{(1)*} \left( \ve{r} \right) \,
      \left\{
      - 
      \frac{\hbar^2}{2m}
      \nabla^2
      +
      V_{\urm{ext}} \left( \ve{r} \right)
      +
      \frac{\delta \tilde{E}_{\urm{Hxc}} \left[ \tilde{\rho}_{\urm{gs}} \right]}{\delta \rho \left( \ve{r} \right)}
      \right\}
      \tilde{\psi}_i \left( \ve{r} \right) 
      \, d \ve{r}
      \notag \\
    & + 
      E_{\urm{Hxc}}^{(1)} \left[ \tilde{\rho}_{\urm{gs}} \right]
      \notag \\
  = & \,
      \sum_{i = 1}^N
      \tilde{\epsilon}_i
      \int
      \tilde{\psi}_i^* \left( \ve{r} \right) \,
      \psi_i^{(1)} \left( \ve{r} \right) 
      \, d \ve{r}
      +
      \sum_{i = 1}^N
      \tilde{\epsilon}_i
      \int
      \psi_i^{(1)*} \left( \ve{r} \right) \,
      \tilde{\psi}_i \left( \ve{r} \right) 
      \, d \ve{r}
      + 
      E_{\urm{Hxc}}^{(1)} \left[ \tilde{\rho}_{\urm{gs}} \right]
      \notag \\
  = & \,
      E_{\urm{Hxc}}^{(1)} \left[ \tilde{\rho}_{\urm{gs}} \right].
\end{align}
It should be noted that the single-particle orbitals $ \tilde{\psi}_i $ are the eigenstates of 
the single-particle Hamiltonian for the non-perturbative system 
\begin{equation}
  \tilde{h}
  =
  - 
  \frac{\hbar^2}{2m}
  \nabla^2
  +
  V_{\urm{ext}} \left( \ve{r} \right)
  +
  \frac{\delta \tilde{E}_{\urm{Hxc}} \left[ \tilde{\rho}_{\urm{gs}} \right]}{\delta \rho \left( \ve{r} \right)}      
\end{equation}
as
\begin{equation}
  \tilde{h} \tilde{\psi}_i = \tilde{\epsilon}_i \tilde{\psi}_i.
\end{equation}
\par
In the latter way, Eq.~\eqref{eq:idea_Hxc} and integration of the KS equation \eqref{eq:KSeq} gives $ E^{\urm{exact}}_{\urm{gs}} $: 
\begin{align}
  E^{\urm{exact}}_{\urm{gs}}
  = & \,
      \sum_{i=1}^N
      \epsilon_i^{\urm{exact}}
      +
      E^{\urm{exact}}_{\urm{Hxc}} \left[ \rho^{\urm{exact}}_{\urm{gs}} \right]
      -
      \int
      \frac{\delta E^{\urm{exact}}_{\urm{Hxc}} \left[ \rho^{\urm{exact}}_{\urm{gs}} \right]}
      {\delta \rho \left(\ve{r} \right)}
      \rho^{\urm{exact}}_{\urm{gs}} \left( \ve{r} \right)
      \, d\ve{r}
      \notag \\
  = & \,
      \sum_{i=1}^N
      \epsilon_i^{\urm{exact}}
      +
      \tilde{E}_{\urm{Hxc}} \left[ \rho^{\urm{exact}}_{\urm{gs}} \right]
      +
      \lambda
      E^{(1)}_{\urm{Hxc}} \left[ \rho^{\urm{exact}}_{\urm{gs}} \right]
      \notag \\
    & -
      \int
      \frac{\delta \tilde{E}_{\urm{Hxc}} \left[ \rho^{\urm{exact}}_{\urm{gs}} \right]}
      {\delta \rho \left(\ve{r} \right)}
      \rho^{\urm{exact}}_{\urm{gs}} \left( \ve{r} \right)
      \, d\ve{r}
      \notag \\
    & -
      \lambda
      \int
      \frac{\delta E^{(1)}_{\urm{Hxc}} \left[ \rho^{\urm{exact}}_{\urm{gs}} \right]}
      {\delta \rho \left(\ve{r} \right)}
      \rho^{\urm{exact}}_{\urm{gs}} \left( \ve{r} \right)
      \, d\ve{r}
      +
      O \left( \lambda^2 \right),
      \label{eq:gsEnergy_2nd}
\end{align}
where $ \epsilon_i^{\urm{exact}} $ are obtained from $ \rho^{\urm{exact}}_{\urm{gs}} $ by using the IKS.
By comparing these two expressions of the ground-state energy
and neglecting $ O \left( \lambda^2 \right) $ term, 
the equation for $ E^{(1)}_{\urm{Hxc}} \left[ \rho \right] $ is obtained:
\begin{align}
  & \lambda E^{(1)}_{\urm{Hxc}} \left[ \tilde{\rho}_{\urm{gs}} \right]
    -
    \lambda E^{(1)}_{\urm{Hxc}} \left[ \rho^{\urm{exact}}_{\urm{gs}} \right]
    +
    \lambda
    \int
    \frac{\delta E^{(1)}_{\urm{Hxc}} \left[ \rho^{\urm{exact}}_{\urm{gs}} \right]}
    {\delta \rho \left( \ve{r} \right)}
    \rho^{\urm{exact}}_{\urm{gs}} \left( \ve{r} \right)
    \, d \ve{r}
    \notag \\
  = & \, 
      \sum_{i=1}^N \epsilon_i^{\urm{exact}}
      +
      \tilde{E}_{\urm{Hxc}} \left[ \rho^{\urm{exact}}_{\urm{gs}} \right]
      -
      \int
      \frac{\delta \tilde{E}_{\urm{Hxc}} \left[ \rho^{\urm{exact}}_{\urm{gs}} \right]}
      {\delta \rho \left( \ve{r} \right)}
      \rho^{\urm{exact}}_{\urm{gs}} \left( \ve{r} \right)
      \, d \ve{r}
      -
      \tilde{E}_{\urm{gs}}
      \notag \\
  \defeq & \,
           C \left[ \rho^{\urm{exact}}_{\urm{gs}} \right].
           \label{eq:basic_eq}
\end{align}
The right-hand side of this equation can be calculated from the known quantities
and
its value depends only on the exact ground-state density $ \rho_{\urm{gs}}^{\urm{exact}} $ 
and the known functional $ \tilde{E}_{\urm{Hxc}} $.
Thus, hereafter the right-hand side of the equation is shown as $ C \left[ \rho \right] $.
\par
Finally, solving Eq.~\eqref{eq:basic_eq},
the Hartree-exchange-correlation functional in the IKS-DFPT in the first-order,
i.e.,~the IKS-DFPT1, is derived as 
\begin{equation}
  \label{eq:calc}
  E_{\urm{Hxc}} \left[ \rho \right]
  =
  \tilde{E}_{\urm{Hxc}} \left[ \rho \right]
  +
  \lambda E^{(1)}_{\urm{Hxc}} \left[ \rho \right].
\end{equation}
\par
Because Eq.~\eqref{eq:basic_eq} is a functional equation, it is difficult to be solved directly.
In this work, we introduce one of the simplest ansatze for $ E_{\urm{Hxc}}^{(1)} \left[ \rho \right] $ within the LDA,
\begin{equation}
  E^{(1)}_{\urm{Hxc}} \left[ \rho \right]
  =
  A
  \int
  \left[
    \rho\left(\ve{r}\right)
  \right]^{\alpha}
  \, d \ve{r},
  \label{eq:PC_new}
\end{equation}
which has the same form as the LDA exchange functional (cf.~Eq.~\eqref{eq:1stCase_EHxc}).
Here, the values of $ A $ and $ \alpha $ are to be determined, and then we get
\begin{equation}
  \label{eq:final_eq}
  \lambda
  A \int
  \left\{
    \left[
      \tilde{\rho}_{\urm{gs}} \left( \ve{r} \right)
    \right]^{\alpha}
    +
    \left( \alpha - 1 \right)
    \left[ \rho^{\urm{exact}}_{\urm{gs}} \left( \ve{r} \right) \right]^{\alpha}
  \right\}
  \, d \ve{r}
  =
  C \left[ \rho^{\urm{exact}}_{\urm{gs}} \right].
\end{equation}
To determine $ A $ and $ \alpha $, two systems, Systems 1 and 2, are required.
Here, $ \rho_1 $ and $ \rho_2 $ are the exact ground-state densities,
and $ \tilde{\rho}_1 $ and $ \tilde{\rho}_2 $ are the ground-state densities calculated from $ \tilde{E}_{\urm{Hxc}} \left[ \rho \right] $ of the Systems 1 and 2, respectively
\footnote{
  If there are $ n $ parameters in the ansatz for $ E^{(1)}_{\urm{Hxc}} $,
  densities for $ n $ systems are required to solve Eq.~\eqref{eq:final_eq}-like algrebraic equation.}.
Substituting $ \rho_i $ and $ \tilde{\rho}_i $ ($ i = 1, \, 2 $) for Eq.~\eqref{eq:basic_eq},
it leads to the two equations for $ \lambda A $ and $ \alpha $.
In such a way, $ \lambda A $ and $ \alpha $ can be determined.
Note that in principle the Hartree-exchange-correlation EDF is system independent,
and thus any system can be used as Systems 1 and 2.
In this paper, the noble-gas atoms are used for two systems
due to simplicity of the spherical symmetry.
\par
\begin{figure}[t]
  \centering
  \includegraphics[width=8.0cm]{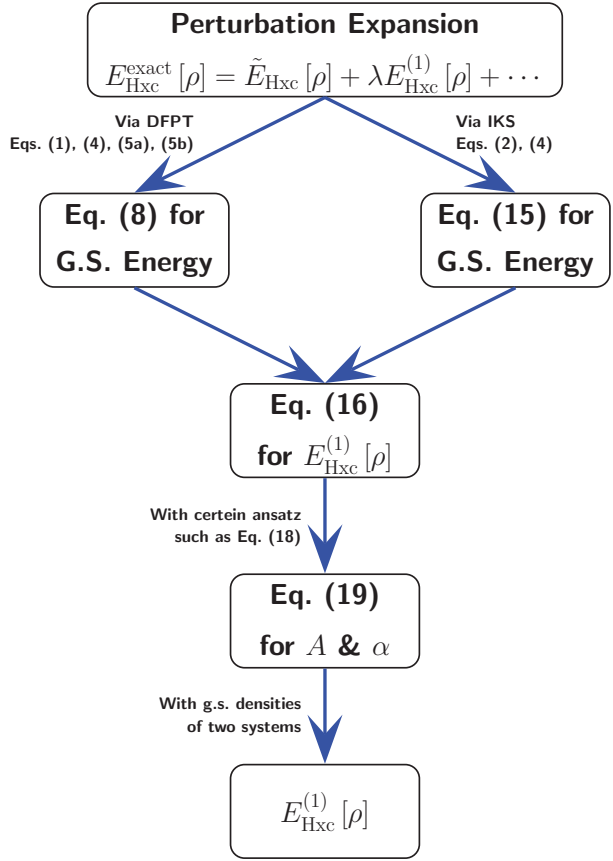}
  \caption{
    Flowchart of the IKS-DFPT method.}
  \label{fig:flowchart}
\end{figure}

In summary, the flowchart of the IKS-DFPT method is shown as Fig.~\ref{fig:flowchart}.
\subsection{Iteration of IKS-DFPT}
\par
On the one hand, a functional derived by the IKS-DFPT1 
$ E_{\urm{Hxc}}^{\urm{1st}} = \tilde{E}_{\urm{Hxc}} + \lambda E_{\urm{Hxc}}^{(1)} $ 
is assumed to be improved from the original functional $ \tilde{E}_{\urm{Hxc}} $.
On the other hand, the functional $ E_{\urm{Hxc}}^{\urm{1st}} $ may still be able to be improved more,
if we repeat the same procedure.
For that, the functional $ E_{\urm{Hxc}}^{\urm{1st}} $ is treated as the conventional functional $ \tilde{E}_{\urm{Hxc}} $ above and applied the IKS-DFPT1 again.
The functional derived by the IKS-DFPT again, $ E_{\urm{Hxc}}^{\urm{2nd}} = E_{\urm{Hxc}}^{\urm{1st}} + \lambda E_{\urm{Hxc}}^{(1)} $, is expected to be improved.
\par
To reach the most improved functional in the IKS-DFPT1 with the ansatz~\eqref{eq:PC_new},
the IKS-DFPT is applied to the derived functional iteratively.
\par
Formally, at the $ n $th step of the iteration,
the Hartree-exchange-correlation functional calculated in the IKS-DFPT1 is
\begin{equation}
  E^{\urm{$ n $th}}_{\urm{Hxc}}\left[ \rho \right]
  =
  \tilde{E}_{\urm{Hxc}}^{\urm{$ 0 $th}} \left[ \rho \right]
  +
  \sum_{k = 1}^n
  \lambda
  E_{\urm{Hxc}}^{\urm{(1), $ k $th}}
\end{equation}
where $ E_{\urm{Hxc}}^{\urm{(1), $ k $th}} $ is the derived term in $ k $th step.
In particular, under the ansatz~\eqref{eq:PC_new},
$ E_{\urm{Hxc}}^{\urm{$ n $th}} \left[ \rho \right] $ is defined as 
\begin{equation}
  E^{\urm{$ n $th}}_{\urm{Hxc}}\left[ \rho \right]
  =
  \tilde{E}_{\urm{Hxc}}^{\urm{$ 0 $th}} \left[ \rho \right]
  +
  \sum_{k = 1}^n
  \lambda 
  A_k
  \int
  \left[
    \rho \left( \ve{r} \right)
  \right]^{\alpha_k}
  \, d \ve{r},
  \label{eq:func_ansatz_it}
\end{equation}
where $ \tilde{E}_{\urm{Hxc}}^{\urm{$ 0 $th}} \left[ \rho \right] $ is the original one
$ \tilde{E}_{\urm{Hxc}} \left[ \rho \right] $ at the first step.
We perform the iteration until convergence.
This indicates that we cannot improve the EDF further by using this scheme and ansatz \eqref{eq:PC_new}.
\section{Benchmark Calculations and Discussions}
\label{sec:calc}
\par
As benchmark calculations,
$ \rho^{\urm{target}}_{\urm{gs}} \left( \ve{r} \right) $ is calculated from the theoretical $ E^{\urm{target}}_{\urm{Hxc}} \left[ \rho \right] $, instead of experimental data,
and we test whether $ E^{\urm{target}}_{\urm{Hxc}} \left[ \rho \right] $ is reproduced in this scheme.
In this section, the superscript ``target'' is used instead of ``exact''
since the functional which should be reproduced is already somehow approximated.
All the pairs of the isolated noble-gas atoms ($ \mathrm{He} $, $ \mathrm{Ne} $, $ \mathrm{Ar} $, $ \mathrm{Kr} $, $ \mathrm{Xe} $, and $ \mathrm{Rn} $) are used as Systems 1 and 2.
In calculations, the Hartree atomic unit is used. 
The ADPACK code \cite{ADPACK} is used for the DFT calculations of the isolated atoms.
Hereafter, $ \lambda A_n $ is denoted by $ A_n $.
\par
We analyze two cases:
1)~$ \tilde{E}_{\urm{Hxc}}^{\urm{$ 0 $th}} \left[ \rho \right] $ is the Hartree functional,
and $ E^{\urm{target}}_{\urm{Hxc}} \left[ \rho \right] $ is the Hartree plus LDA exchange functional \cite{Proc.Camb.Phil.Soc.26.376}:
\numparts
\begin{align}
  \tilde{E}_{\urm{Hxc}}^{\urm{$ 0 $th}} \left[ \rho \right]
  & =
    \frac{1}{2}
    \iint
    \frac{\rho \left( \ve{r} \right) \, \rho \left( \ve{r}' \right)}
    {\left| \ve{r} - \ve{r}' \right|}
    \, d \ve{r}
    \, d \ve{r}', 
    \label{eq:1stCase_tilde_EHxc} \\
  E^{\urm{target}}_{\urm{Hxc}} \left[ \rho \right]
  & =
    \tilde{E}_{\urm{Hxc}}^{\urm{$ 0 $th}} \left[ \rho \right]
    -
    \frac{3}{4}
    \left(
    \frac{3}{\pi}
    \right)^{1/3}
    \int
    \left[
    \rho \left( \ve{r} \right)
    \right]^{4/3}
    \, d \ve{r}.
    \label{eq:1stCase_EHxc}
\end{align}
\endnumparts
2)~$ \tilde{E}_{\urm{Hxc}}^{\urm{$ 0 $th}} \left[ \rho \right] $ is the Hartree plus LDA exchange functional,
and $ E^{\urm{target}}_{\urm{Hxc}} \left[ \rho \right] $ is the Hartree plus LDA exchange-correlation functional,
where the PZ81 \cite{PhysRevB.23.5048} functional $ E_{\urm{c}}^{\urm{PZ81}} \left[ \rho \right] $ is used:
\numparts
\begin{align}
  \tilde{E}_{\urm{Hxc}}^{\urm{$ 0 $th}} \left[ \rho \right]
  & = 
    \frac{1}{2}
    \iint
    \frac{\rho \left( \ve{r} \right) \, \rho \left( \ve{r} '\right)}
    {\left| \ve{r} - \ve{r}' \right|}
    \, d \ve{r}
    \, d \ve{r}' 
    - 
    \frac{3}{4}
    \left(
    \frac{3}{\pi}
    \right)^{1/3}
    \int
    \left[
    \rho \left( \ve{r} \right)
    \right]^{4/3}
    \, d \ve{r}, 
    \label{eq:2ndCase_tilde_EHxc} \\
  E^{\urm{target}}_{\urm{Hxc}} \left[ \rho \right]
  & = 
    \tilde{E}_{\urm{Hxc}}^{\urm{$ 0 $th}} \left[ \rho \right]
    +
    E_{\urm{c}}^{\urm{PZ81}} \left[ \rho \right].
    \label{eq:2ndCase_EHxc}
\end{align}
\endnumparts
In both cases, the external field
$ V^{\urm{target}}_{\urm{ext}} \left( \ve{r} \right) = \tilde{V}_{\urm{ext}} \left( \ve{r} \right) $
is the Coulomb interaction between the nucleus and electron.
\par
First, let us consider the first case, i.e.,~the calculations
from the Hartree approximation \eqref{eq:1stCase_tilde_EHxc} to the Hartree--Fock--Slater (LDA exchange) approximation \eqref{eq:1stCase_EHxc}.
In Tables~\ref{tab:it_hx_HeNe} and \ref{tab:it_hx_XeRn},
the coefficients $ \lambda A_n $ and $ \alpha_n $ and the ground-state energies $ E_{\urm{gs}}^{\urm{$ n $th}} $ calculated in the $ n $th iteration are shown for the pairs of atoms $ \mathrm{He} $-$ \mathrm{Ne} $ and $ \mathrm{Xe} $-$ \mathrm{Rn} $, respectively.
It is found that $ \lambda A_1 $ and $ \alpha_1 $ are obtained within $ 7.2 \, \% $ and $ 1.0 \, \% $ errors in $ \mathrm{He} $-$ \mathrm{Ne} $,
and within $ 2.3 \, \% $ and $ 0.2 \, \% $ errors in $ \mathrm{Xe} $-$ \mathrm{Rn} $, respectively, from their target values.
The heavier atoms reproduce the coefficients better.
The results of the other pairs are shown in the Appendix.
\par
\begin{figure}[t]
  \centering
  \includegraphics[width=8cm]{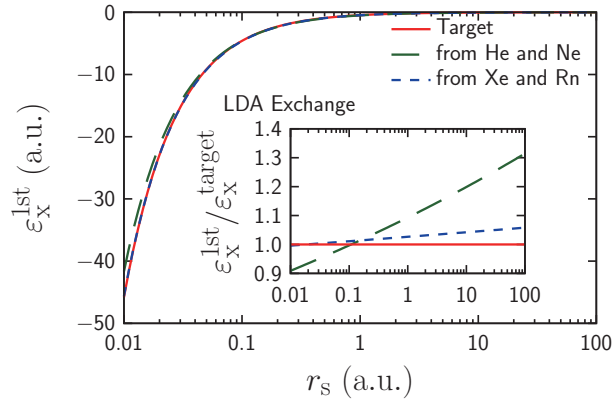}
  \caption{
    Energy density $ \epsilon^{\urm{$ 1 $st}}_{\urm{x}} $ for the LDA exchange functional calculated in the first iteration as functions of $ r_{\urm{s}} $.
    Results for the pairs of $ \mathrm{He} $-$ \mathrm{Ne} $ 
    and 
    $ \mathrm{Xe} $-$ \mathrm{Rn} $ are shown with dashed and dot lines, respectively.
    The target one is shown with a solid line.
    Ratios of $ \epsilon^{\urm{$ 1 $st}}_{\urm{x}} / \epsilon^{\urm{target}}_{\urm{x}} $ are shown in the insert.}
  \label{fig:hx_HeNe_XeRn}
\end{figure}

The exchange energy density calculated in the first iteration,
$ \epsilon^{\urm{$ 1 $st}}_{\urm{x}} \left( r_{\urm{s}} \right) $,
and the ratio to the target one,
$ \epsilon^{\urm{$ 1 $st}}_{\urm{x}} \left( r_{\urm{s}} \right) / \epsilon^{\urm{target}}_{\urm{x}} \left( r_{\urm{s}} \right) $,
are shown as functions of $ r_{\urm{s}} $ in Fig.~\ref{fig:hx_HeNe_XeRn}
for the pairs of $ \mathrm{He} $-$ \mathrm{Ne} $ and $ \mathrm{Xe} $-$ \mathrm{Rn} $ with dashed and dot lines, respectively,
while the target one is shown with a solid line.
Here, the energy density $ \epsilon_i \left(\rho\right) $
and the Wigner-Seitz radius $ r_{\urm{s}} $ are defined as
$ E_i \left[ \rho \right] =
\int
\epsilon_i \left( \rho \right) \,
\rho \left( \ve{r} \right)
\, d \ve{r}
$ ($ i = \mathrm{x} $, $ \mathrm{c} $) and 
$ r_{\urm{s}} =
\left[
  3 / \left(4 \pi \rho \right)
\right]^{1/3} $, respectively.
The pair of $ \mathrm{Xe} $-$ \mathrm{Rn} $ reproduces the target functional within a few percents in the range of $ 0.01 \, \mathrm{a.u.} \le r_{\urm{s}} \le 100 \, \mathrm{a.u.} $,
which is generally better than the pair of $ \mathrm{He} $-$ \mathrm{Ne} $.
As comparing to
$ \mathrm{He} $-$ \mathrm{Ne} $ and $ \mathrm{Xe} $-$ \mathrm{Rn} $ cases,
since the polynomial form of the functional in Eq.~\eqref{eq:PC_new} is more sensitive to the high-density region,
better reproduction in the high-density region leads to better reproduction of the coefficients.
\par
For the iterations,
it is found in Tables \ref{tab:it_hx_HeNe} and \ref{tab:it_hx_XeRn} that the difference between $ E_{\urm{gs}}^{\urm{$ n $th}} $ and the target $ E^{\urm{target}}_{\urm{gs}} $ becomes smaller as the iteration proceeds further.
The ground-state energies of $ \mathrm{He} $, $ \mathrm{Ne} $, $ \mathrm{Xe} $, and $ \mathrm{Rn} $ are finally reproduced within $ 0.4 \, \% $, $ 0.003 \, \% $, $ 0.002 \, \% $, and $ 0.0003 \, \% $ errors, respectively,
comparing with $ 28 \, \% $, $ 8 \, \% $, $ 2 \, \% $, and $ 2 \, \% $ errors at the zeroth step, $ E_{\urm{gs}}^{\urm{$ 0 $th}} $.
This indicates the iteration helps the improvement of the ground-state energy.
\par
\begin{figure}[t]
  \centering
  \includegraphics[width=8cm]{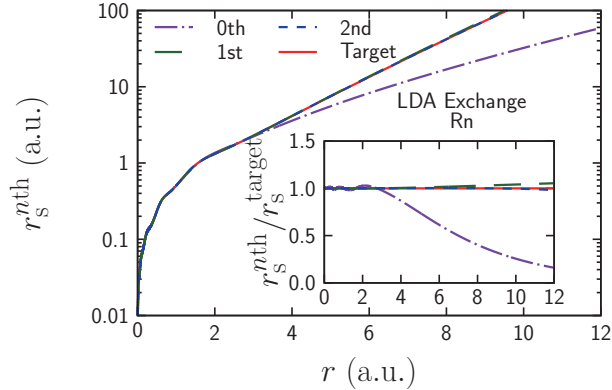}
  \caption{
    Wigner-Seitz radii $ r^{\urm{$ n $th}}_{\urm{s}} $ as functions of $ r $ for $ \mathrm{Rn} $.
    Results calculated in the zeroth, first, and second iterations are shown with dot-dashed, dashed, and dot lines, respectively.
    The target one is shown as a solid line.
    Ratios of $ r^{\urm{$ n $th}}_{\urm{s}} / r^{\urm{target}}_{\urm{s}} $ are shown in the insert.}
  \label{fig:it_density_Rn_hx}
\end{figure}

The Wigner-Seitz radii $ r^{\urm{$ n $th}}_{\urm{s}} $ calculated in the zeroth, first, and second iterations
and the target one $ r_{\urm{s}}^{\urm{target}} $ for $ \mathrm{Rn} $ 
are shown as functions of $ r $ in Fig.~\ref{fig:it_density_Rn_hx}
with dot-dashed, dashed, dot, and solid lines, respectively.
The ratio of calculated Wigner-Seitz radius to the target one, 
$ r^{\urm{$ n $th}}_{\urm{s}} / r^{\urm{target}}_{\urm{s}} $,
for each step is also shown in the insert of Fig.~\ref{fig:it_density_Rn_hx}.
It is found that the ground-state density at the first step is already much improved,
and it is even further improved as the iteration proceeds further.
This indicates the iteration also helps the improvement of the ground-state density.
\begin{table}[t]
  \caption{
    The coefficients $ \lambda A_n $ and $ \alpha_n $ and the ground-state energies $ E_{\urm{gs}}^{\urm{$ n $th}} $ calculated in
    the $ n $th iteration
    shown in Eq.~\eqref{eq:func_ansatz_it}
    for the pair of atoms $ \mathrm{He} $ and $ \mathrm{Ne} $.
    The Hartree functional given in Eq.~\eqref{eq:1stCase_tilde_EHxc} is used for
    $ \tilde{E}^{\urm{$ 0 $th}}_{\urm{Hxc}}\left[\rho\right] $
    and the Hartree plus LDA exchange functional given in 
    Eq.~\eqref{eq:1stCase_EHxc} are used for $ E^{\urm{target}}_{\urm{Hxc}} \left[ \rho \right] $.
    All units are in the Hartree atomic unit.}
  \label{tab:it_hx_HeNe}
  \centering
  \begin{tabular}{crrrr}
    \hline \hline
    $ n $ & \multicolumn{1}{c}{$ \alpha_n $} & \multicolumn{1}{c}{$ \lambda A_n $} & \multicolumn{1}{c}{$ E^{\urm{$ n $th}}_{\urm{gs}}$ of $ \mathrm{He} $} & \multicolumn{1}{c}{$ E^{\urm{$ n $th}}_{\urm{gs}}$ of $ \mathrm{Ne} $} \\ \hline
    $ 0 $ & & & $ -1.9517070 $ & $ -116.99029 $ \\ 
    $ 1 $ & $ 1.3199872 $ & $ -0.7920448 $ & $ -2.8010654 $ & $ -127.95544 $ \\
    $ 2 $ & $ 1.0125782 $ & $ 0.0459594 $ & $ -2.7115372 $ & $ -127.49446 $ \\ \hline
    Target & $ 1.3333333 $ & $ -0.7385588 $ & $ -2.7237069 $ & $ -127.49109 $ \\
    \hline \hline
  \end{tabular}
\end{table}
\begin{table}[t]
  \caption{
    Same as Table \ref{tab:it_hx_HeNe} but for the pair of atoms $ \mathrm{Xe} $ and $ \mathrm{Rn} $.}
  \label{tab:it_hx_XeRn}
  \centering
  \begin{tabular}{crrrr}
    \hline \hline
    $ n $ & \multicolumn{1}{c}{$ \alpha_n $} & \multicolumn{1}{c}{$ \lambda A_n $} & \multicolumn{1}{c}{$ E^{\urm{$ n $th}}_{\urm{gs}}$ of $ \mathrm{Xe} $} & \multicolumn{1}{c}{$ E^{\urm{$ n $th}}_{\urm{gs}}$ of $ \mathrm{Rn} $} \\ \hline
    $ 0 $ & & & $ -7054.6485 $ & $ -21479.344 $ \\
    $ 1 $ & $ 1.3311445 $ & $ -0.7558229 $ & $ -7224.9365 $ & $ -21852.010 $ \\
    $ 2 $ & $ 1.0436323 $ & $ 0.0306234 $ & $ -7223.0601 $ & $ -21848.894 $ \\ \hline
    Target & $ 1.3333333 $ & $ -0.7385588 $ & $ -7223.1853 $ & $ -21848.954 $ \\ 
    \hline \hline
  \end{tabular}
\end{table}
%
\par
Next, let us consider the second case, i.e.,~the calculations
from the Hartree--Fock--Slater (LDA exchange) approximation \eqref{eq:2ndCase_tilde_EHxc} to the LDA exchange-correlation \eqref{eq:2ndCase_EHxc}.
In Tables \ref{tab:it_hxc_HeNe} and \ref{tab:it_hxc_XeRn},
the coefficients $ \lambda A_n $ and $ \alpha_n $ and the ground-state energies $ E^{\urm{$ n $th}}_{\urm{gs}}$
calculated in the $ n $th iteration are shown for the pairs of atoms $ \mathrm{He} $-$ \mathrm{Ne} $ and $ \mathrm{Xe} $-$ \mathrm{Rn} $, respectively.
It is found that the ground-state energies are already reproduced well at the first step.
Here, for the pair of $ \mathrm{Xe} $-$ \mathrm{Rn} $, the convergence reaches.
For the pair of $ \mathrm{He} $-$ \mathrm{Ne} $, the ground-state energies are further improved slightly as the iteration proceeds further.
The ground-state energies of $ \mathrm{He} $, $ \mathrm{Ne} $, $ \mathrm{Xe} $, and $ \mathrm{Rn} $ are finally reproduced within $ 0.07 \, \% $, $ 0.0009 \, \% $, $ 0.0005 \, \% $, and $ 0.0001 \, \% $ error, respectively,
comparing with $ 4 \, \% $, $ 0.6 \, \% $, $ 0.07 \, \% $, and $ 0.04 \, \% $ errors at the zeroth step, $ E_{\urm{gs}}^{\urm{$ 0 $th}} $.
The results of the other pairs are shown in the Appendix.
\par
\begin{figure}[t]
  \centering
  \includegraphics[width=8cm]{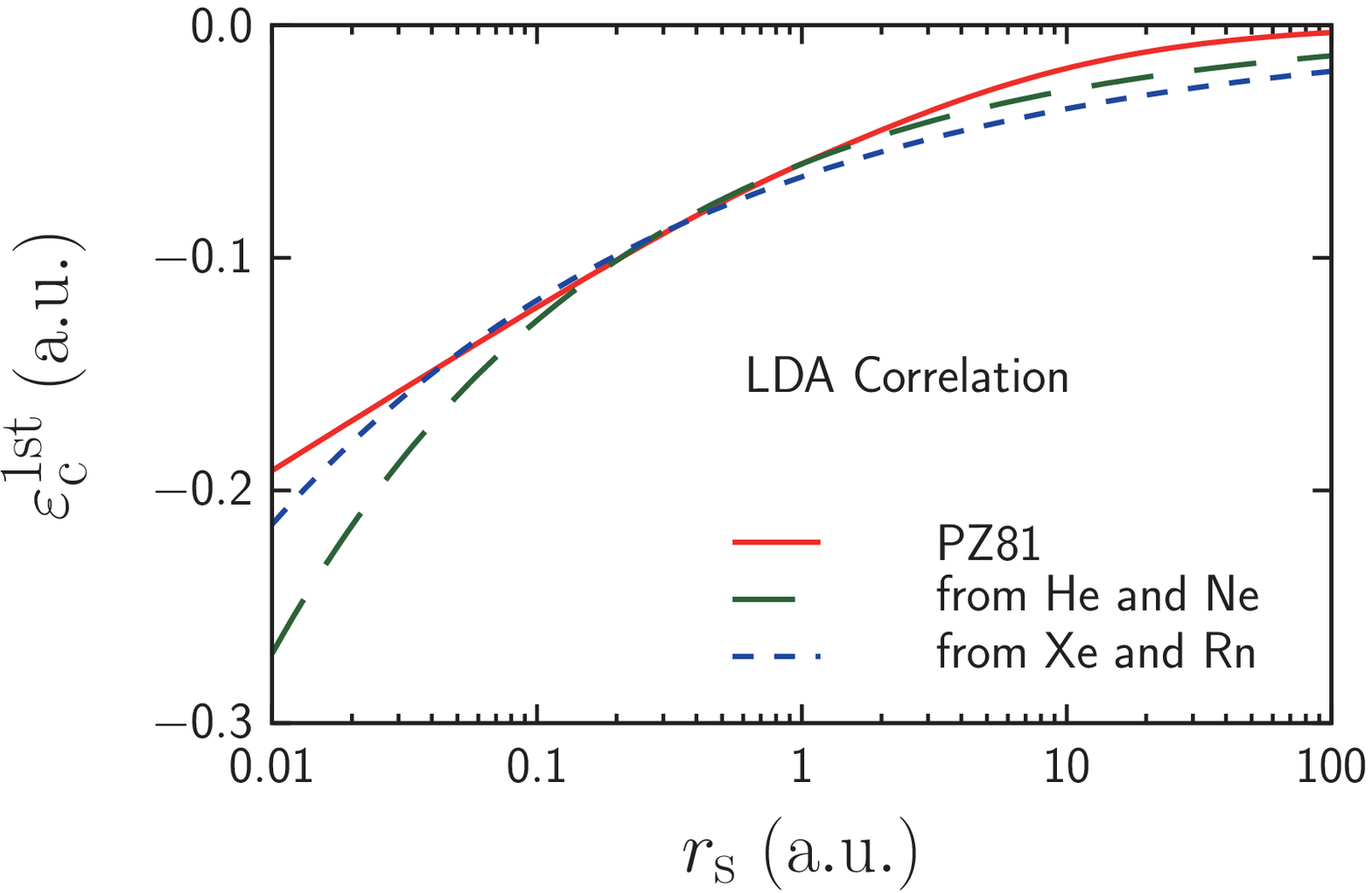}
  \caption{
    Same as Fig.~\ref{fig:hx_HeNe_XeRn} but $ \epsilon^{\urm{$ 1 $st}}_{\urm{c}} $ for the LDA correlation functional.}
  \label{fig:hxc_HeNe_XeRn}
\end{figure}

In order to compare the calculated correlation functionals and the target one,
the correlation energy density calculated in the first iteration
$ \epsilon^{\urm{$ 1 $st}}_{\urm{c}} \left( r_{\urm{s}} \right) $
is shown as functions of $ r_{\urm{s}} $ in Fig.~\ref{fig:hxc_HeNe_XeRn}
for the pairs of $ \mathrm{He} $-$ \mathrm{Ne} $ and $ \mathrm{Xe} $-$ \mathrm{Rn} $ with dashed and dot lines, respectively,
while the target one is shown with a solid line.
The non-polynomial PZ81 functional is reproduced better in the lower-density region from the pair of $ \mathrm{He} $-$ \mathrm{Ne} $, and in the higher-density region from the pair of $ \mathrm{Xe} $-$ \mathrm{Rn} $,
since heavier atoms have higher-density region than lighter atoms.
\par
\begin{figure}[t]
  \centering
  \includegraphics[width=8cm]{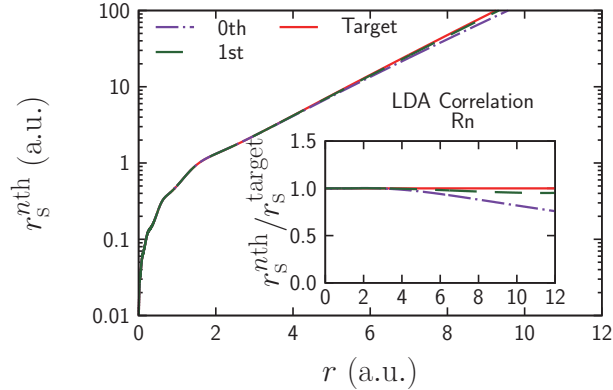}
  \caption{
    Wigner-Seitz radii $ r^{\urm{$ n $th}}_{\urm{s}} $ as functions of $ r $ for $ \mathrm{Rn} $.
    Results calculated in the zeroth and first iterations are shown with dot-dashed and dashed lines, respectively.
    The target one is shown as a solid line.
    Ratios of $ r^{\urm{$ n $th}}_{\urm{s}} / r^{\urm{target}}_{\urm{s}} $ are shown in the insert.}
  \label{fig:it_density_Rn_hxc}
\end{figure}

The Wigner-Seitz radii $ r^{\urm{$ n $th}}_{\urm{s}} $ calculated in the zeroth and first iterations
and the target one $ r_{\urm{s}}^{\urm{target}} $ for $ \mathrm{Rn} $ 
are shown as functions of $ r $ in Fig.~\ref{fig:it_density_Rn_hxc}
with dot-dashed, dashed, and solid lines, respectively.
The ratio of calculated Wigner-Seitz radius to the target one, 
$ r^{\urm{$ n $th}}_{\urm{s}} / r^{\urm{target}}_{\urm{s}} $,
for each step is also shown in the insert of Fig.~\ref{fig:it_density_Rn_hxc}.
It is found that the ground-state density at the first step is already much improved.
\par
Comparing with the above two cases,
in the first case, we note that the difference between $ \tilde{E}_{\urm{Hxc}}^{\urm{$ 0 $th}} \left[ \rho \right] $ and $ E_{\urm{Hxc}}^{\urm{target}} \left[ \rho \right] $ is larger,
and thus either more iteration steps are required,
or we should consider the IKS with the second-order DFPT.
Nevertheless, as the LDA exchange functional is polynomial,
the energy density $ \epsilon_{\urm{x}} $ is reproduced well in wide-density range.
In the second case, the difference between $ \tilde{E}_{\urm{Hxc}}^{\urm{$ 0 $th}} \left[ \rho \right] $ and $ E_{\urm{Hxc}}^{\urm{target}} \left[ \rho \right] $ is smaller,
and thus the IKS-DFPT1 is enough to reproduce the ground-state energy and density.
However, as the PZ81 functional is non-polynomial, the energy density $ \epsilon_{\urm{c}} $ is not reproduced well in wide-density range within the present polynomial ansatz.
We should go beyond the polynomial ansatz in the future.
\begin{table}[t]
  \caption{
    Same as Table~\ref{tab:it_hx_HeNe} but
    the Hartree plus LDA exchange functional given in Eq.~\eqref{eq:2ndCase_tilde_EHxc} is used for $ \tilde{E}_{\urm{Hxc}}^{\urm{$ 0 $th}} \left[ \rho \right] $ 
    and the Hartree plus LDA exchange-correlation functional given in Eq.~\eqref{eq:2ndCase_EHxc}
    is used for $ E^{\urm{target}}_{\urm{Hxc}} \left[ \rho \right] $.}
  \label{tab:it_hxc_HeNe}
  \centering
  \begin{tabular}{crrrr}
    \hline \hline
    $ n $ & \multicolumn{1}{c}{$ \alpha_n $} & \multicolumn{1}{c}{$ \lambda A_n $} & \multicolumn{1}{c}{$ E^{\urm{$ n $th}}_{\urm{gs}}$ of $ \mathrm{He} $} & \multicolumn{1}{c}{$ E^{\urm{$ n $th}}_{\urm{gs}}$ of $ \mathrm{Ne} $} \\ \hline
    $ 0 $ & & & $ -2.7237069 $ & $ -127.49109 $ \\ 
    $ 1 $ & $ 1.1093168 $ & $ -0.0697561 $ & $ -2.8365158 $ & $ -128.22944 $ \\
    $ 2 $ & $ 0.6951714 $ & $ 0.0000544 $ & $ -2.8362598 $ & $ -128.22881 $ \\ \hline
    Target & \multicolumn{1}{c}{PZ81} & \multicolumn{1}{c}{PZ81} & $ -2.8343506 $ & $ -128.22766 $ \\ 
    \hline \hline
  \end{tabular}
\end{table}
\begin{table}[t]
  \caption{
    Same as Table \ref{tab:it_hxc_HeNe} but for the pair of atoms $ \mathrm{Xe} $ and $ \mathrm{Rn} $.}
  \label{tab:it_hxc_XeRn}
  \centering
  \begin{tabular}{crrrr}
    \hline \hline
    $ n $ & \multicolumn{1}{c}{$ \alpha_n $} & \multicolumn{1}{c}{$ \lambda A_n $} & \multicolumn{1}{c}{$ E^{\urm{$ n $th}}_{\urm{gs}}$ of $ \mathrm{Xe} $} & \multicolumn{1}{c}{$ E^{\urm{$ n $th}}_{\urm{gs}}$ of $ \mathrm{Rn} $} \\ \hline
    $ 0 $ & & & $ -7223.1853 $ & $ -21848.954 $ \\ 
    $ 1 $ & $ 1.0862074 $ & $ -0.0737520 $ & $ -7228.4020 $ & $ -21857.981 $ \\ \hline
    Target & \multicolumn{1}{c}{PZ81} & \multicolumn{1}{c}{PZ81} & $ -7228.3628 $ & $ -21857.954 $ \\ 
    \hline \hline
  \end{tabular}
\end{table}
%
\section{Conclusion and Perspectives}
\label{sec:conc}
\par
In summary, the way to improve conventional EDFs based on the combination of the IKS and the DFPT is proposed.
As benchmark calculations, we test whether the LDA exchange and correlation functionals can be reproduced in this novel scheme IKS-DFPT1.
It is found that with the present polynomial ansatz the polynomial functional can be well reproduced, 
while the non-polynomial one can be reproduced in the crucial density region.
By improving the exchange and correlation functionals, the accuracy of the ground-state energies is improved by two to three orders of magnitude,
and the accuracy of the ground-state densities is also improved one to two orders of magnitude.
Therefore, the IKS-DFPT is promising to improve the conventional functionals.
\par
In this paper, we mainly focus on the feasibility of this method by using  
a simple polynomial ansatz shown in Eq.~\eqref{eq:PC_new}
and the noble-gas atoms.
To get more accurate functional, beyond polynomial ansatz of $ E_{\urm{Hxc}}^{(1)} \left[ \rho \right] $,
including considering the density gradient,
is one of the interesting topics.
Also, in this paper, two combinations of two systems,
$ \mathrm{He} $-$ \mathrm{Ne} $ and $ \mathrm{Xe} $-$ \mathrm{Rn} $,
are used as a benchmark,
while the results for the other pairs are shown in the Appendix.
It is a future task to get optimized values across all possible pairs or among the essentially different systems.
As another perspective, the second-order IKS-DFPT is interesting.
It is also important to include the spin and isospin degrees of freedom for applications of spin-polarized electron systems or nuclear systems.
\par
According to the Hohenberg--Kohn theorem, 
the functional is, in principle, system independent.
In practice, due to approximations, such as the first-order perturbation theory and the ansatz \eqref{eq:PC_new},
$ \alpha_n $ and $ \lambda A_n $ have slightly system dependence as shown in the Appendix.
If these approximations are appropriate, the system dependence is quite small.
Therefore, any system can be used for this purpose as long as the density is known,
and 
once densities of several systems are known
the EDFs can be improved by using this novel method IKS-DFPT for actual problems.
%
\ack
\par
The authors appreciate
Ryosuke Akashi, Gianluca Col\`{o}, Yixin Guo, 
Xavier Roca-Maza, Osamu Sugino, Shinji Tsuneyuki, and Dario Vretenar
for stimulating discussions and valuable comments.
T.N.~and D.O.~acknowledge the financial support from Computational Science Alliance, The University of Tokyo.
T.N.~and H.L.~would like to thank the RIKEN iTHEMS program
and the JSPS-NSFC Bilateral Program for Joint Research Project on Nuclear mass and life for unravelling mysteries of the $ r $-process.
T.N.~acknowledges the JSPS Grant-in-Aid for JSPS Fellows under Grant No.~19J20543.
H.L.~acknowledges the JSPS Grant-in-Aid for Early-Career Scientists under Grant No.~18K13549.
\appendix
\section{Results for All Pairs of Noble Gases}
\par
The coefficients $ \lambda A_1 $ and $ \alpha_1 $ calculated in 
the first iteration for all the pairs of noble-gas atoms are shown in Table \ref{tab:App_hx_all}.
\begin{table*}[!htb]
  \caption{
    The coefficients $ \lambda A_1 $ and $ \alpha_1 $ calculated in 
    the first iteration for all the pairs of noble-gas atoms.
    All units are in the Hartree atomic unit.}
  \label{tab:App_hx_all}
  \centering
  \begin{tabular}{lp{8em}p{8em}p{8em}p{8em}}
    \hline \hline
    \multicolumn{1}{c}{Pairs} & \multicolumn{1}{c}{Exchange $ \alpha_1 $} & \multicolumn{1}{c}{Exchange $ \lambda A_1 $} & \multicolumn{1}{c}{Correlation $ \alpha_1 $} & \multicolumn{1}{c}{Correlation $ \lambda A_1 $} \\ \hline
    $ \mathrm{He} $ and $ \mathrm{Ne} $ & \multicolumn{1}{c}{$ 1.3199872 $} & \multicolumn{1}{c}{$ -0.7920448 $} & \multicolumn{1}{c}{$ 1.1093169 $} & \multicolumn{1}{c}{$ -0.0697561 $} \\
    $ \mathrm{He} $ and $ \mathrm{Ar} $ & \multicolumn{1}{c}{$ 1.3209765 $} & \multicolumn{1}{c}{$ -0.7926638 $} & \multicolumn{1}{c}{$ 1.1049002 $} & \multicolumn{1}{c}{$ -0.0694846 $} \\
    $ \mathrm{Ne} $ and $ \mathrm{Ar} $ & \multicolumn{1}{c}{$ 1.3235352 $} & \multicolumn{1}{c}{$ -0.7841588 $} & \multicolumn{1}{c}{$ 1.0916279 $} & \multicolumn{1}{c}{$ -0.0718811 $} \\
    $ \mathrm{He} $ and $ \mathrm{Kr} $ & \multicolumn{1}{c}{$ 1.3227758 $} & \multicolumn{1}{c}{$ -0.7937863 $} & \multicolumn{1}{c}{$ 1.1022591 $} & \multicolumn{1}{c}{$ -0.0693213 $} \\
    $ \mathrm{Ne} $ and $ \mathrm{Kr} $ & \multicolumn{1}{c}{$ 1.3263436 $} & \multicolumn{1}{c}{$ -0.7779323 $} & \multicolumn{1}{c}{$ 1.0941254 $} & \multicolumn{1}{c}{$ -0.0715841 $} \\ 
    $ \mathrm{Ar} $ and $ \mathrm{Kr} $ & \multicolumn{1}{c}{$ 1.3290958 $} & \multicolumn{1}{c}{$ -0.7658732 $} & \multicolumn{1}{c}{$ 1.0956532 $} & \multicolumn{1}{c}{$ -0.0711566 $} \\
    $ \mathrm{He} $ and $ \mathrm{Xe} $ & \multicolumn{1}{c}{$ 1.3235844 $} & \multicolumn{1}{c}{$ -0.7942892 $} & \multicolumn{1}{c}{$ 1.1001295 $} & \multicolumn{1}{c}{$ -0.0691892 $} \\
    $ \mathrm{Ne} $ and $ \mathrm{Xe} $ & \multicolumn{1}{c}{$ 1.3270817 $} & \multicolumn{1}{c}{$ -0.7762984 $} & \multicolumn{1}{c}{$ 1.0919878 $} & \multicolumn{1}{c}{$ -0.0718384 $} \\
    $ \mathrm{Ar} $ and $ \mathrm{Xe} $ & \multicolumn{1}{c}{$ 1.3292187 $} & \multicolumn{1}{c}{$ -0.7654719 $} & \multicolumn{1}{c}{$ 1.0921353 $} & \multicolumn{1}{c}{$ -0.0717899 $} \\
    $ \mathrm{Kr} $ and $ \mathrm{Xe} $ & \multicolumn{1}{c}{$ 1.3294148 $} & \multicolumn{1}{c}{$ -0.7644846 $} & \multicolumn{1}{c}{$ 1.0848635 $} & \multicolumn{1}{c}{$ -0.0742007 $} \\
    $ \mathrm{He} $ and $ \mathrm{Rn} $ & \multicolumn{1}{c}{$ 1.3244450 $} & \multicolumn{1}{c}{$ -0.7948236 $} & \multicolumn{1}{c}{$ 1.0980694 $} & \multicolumn{1}{c}{$ -0.0690610 $} \\
    $ \mathrm{Ne} $ and $ \mathrm{Rn} $ & \multicolumn{1}{c}{$ 1.3279028 $} & \multicolumn{1}{c}{$ -0.7744818 $} & \multicolumn{1}{c}{$ 1.0905537 $} & \multicolumn{1}{c}{$ -0.0720085 $} \\
    $ \mathrm{Ar} $ and $ \mathrm{Rn} $ & \multicolumn{1}{c}{$ 1.3297748 $} & \multicolumn{1}{c}{$ -0.7636589 $} & \multicolumn{1}{c}{$ 1.0902537 $} & \multicolumn{1}{c}{$ -0.0721279 $} \\
    $ \mathrm{Kr} $ and $ \mathrm{Rn} $ & \multicolumn{1}{c}{$ 1.3303022 $} & \multicolumn{1}{c}{$ -0.7606336 $} & \multicolumn{1}{c}{$ 1.0856559 $} & \multicolumn{1}{c}{$ -0.0739752 $} \\
    $ \mathrm{Xe} $ and $ \mathrm{Rn} $ & \multicolumn{1}{c}{$ 1.3311445 $} & \multicolumn{1}{c}{$ -0.7558229 $} & \multicolumn{1}{c}{$ 1.0862074 $} & \multicolumn{1}{c}{$ -0.0737520 $} \\ \hline 
    Target & \multicolumn{1}{c}{$ 1.3333333 $} & \multicolumn{1}{c}{$ -0.7385588 $} & \multicolumn{1}{c}{PZ81} & \multicolumn{1}{c}{PZ81} \\ \hline \hline
  \end{tabular}
\end{table*}
\clearpage
%
\providecommand{\newblock}{}

\end{document}